# A comprehensive diagram to grow InAlN alloys by plasma-assisted molecular beam epitaxy

S. Fernández-Garrido[*], Ž. Gačević, and E. Calleja

*ISOM and Dpt. de Ingeniería Electrónica, ETSI Telecomunicación*

*Universidad Politécnica de Madrid, 28040 Madrid, Spain*

Indium incorporation and surface morphology of InAlN layers grown on (0001) GaN by plasma–assisted molecular beam epitaxy were investigated as a function of the impinging In flux and the substrate temperature in the 450-610ºC range. In incorporation was found to decrease with substrate temperature due to thermal decomposition of the growing layer, while for a given temperature it increased with the impinging In flux until stoichiometry was reached at the growth front. The InN losses during growth followed an Arrhenius behaviour characterized by an activation energy of 2.0 eV. A growth diagram highly instrumental to identify optimum growth conditions was established.

---

[*] electronic mail: sfernandez@die.upm.es

InAlN alloys have a direct band gap tunable from 0.7 eV to 6.2 eV, and, for a 17% In there is lattice match (in-plane) to GaN. Heterostrutures and devices including InAlN layers, such as resonant cavities, multi-quantum wells for high-speed inter-subband devices, or high electron mobility transistors have been recently reported [1-4].

Growth diagrams, useful to identify optimum conditions, have been established for binaries (AlN, GaN, InN), grown by plasma-assisted molecular beam epitaxy (PA-MBE), from the surface morphology dependence on growth temperature and impinging fluxes [5-8]. However, the growth of InAl(Ga)N alloys poses much more difficulties due to InN thermal decomposition [9-12] and strong differences between InN and AlN.

This work reports on the growth and characterization of metal-face InAlN layers on GaN templates. Indium incorporation and surface morphology are analysed as a function of growth temperature and metal fluxes to build up a growth diagram.

InAlN thin layers (~80 nm thick) were grown by PA-MBE on (0001) GaN templates (~3.6 µm thick) grown by metal-organic vapour phase epitaxy on sapphire (Lumilog). Growth temperature was measured with an Ircon Modline3 optical pyrometer. Metal fluxes ($\Phi_{Ga}$, $\Phi_{Al}$, $\Phi_{In}$), measured as Beam Equivalent Pressure (Bayard Alpert), were calibrated in atoms/sec·cm$^2$ using cross-sectional scanning electron microscopy (SEM) data from N-rich GaN, AlN and InN thick layers grown at temperatures where thermal decomposition and adatoms desorption are negligible [5]. Similarly, the active N flux, $\Phi_N$, was calibrated in atoms/sec·cm$^2$ using cross-sectional SEM data from Ga-rich GaN thick layers grown at low temperatures (680 ºC). Prior to the InAlN growth, a 100 nm thick GaN buffer layer was grown at 700 ºC under intermediate Ga-rich conditions [5] to obtain a smooth and flat surface. Alloy compositions were assessed by high resolution x-ray diffraction (HR-XRD) and surface morphologies were characterized by SEM (JEOL JSM-5800) and by atomic force microscopy (AFM, Digital Instruments MMAFM-2).

To analyse separately the effects of growth temperature and impinging In flux on In incorporation two sets of samples were grown. In a first set (Series-A) all impinging fluxes were kept constant, and the growth temperature varied between 450 and 610ºC, a range where Al sticking coefficient is 1, being preferentially incorporated due to a much higher energy of the Al-N bond than that of In-N [9,11]. The III/V flux ratio was <1 (*N-rich*) with impinging fluxes of $\Phi_{Al} = 2.7 \pm 0.1 \times 10^{14}$ atoms/s·cm², $\Phi_{In} = 1.2 \pm 0.1 \times 10^{14}$ atoms/s·cm² and $\Phi_{N} = 4.2 \pm 0.1 \times 10^{14}$ atoms/s·cm². A $0.31 \pm 0.03$ nominal InN mole fraction is derived from $\Phi_{In}/(\Phi_{In}+\Phi_{Al})$ [9]. In Series-B the growth temperature was set at 535 ºC and $\Phi_{In}$ was varied while keeping the same $\Phi_{Al}$ and $\Phi_{N}$ of Series-A.

Figure 1(a) shows a continuous decrease of the *actual* InN mole fraction, $[InN]^*$, (Series-A) with the growth temperature, from the nominal value (considering the experimental error derived from the calibration of the impinging fluxes), 0.33 @ 450 ºC down to 0.02 @ 607 ºC. For temperatures bellow 500 ºC In droplets were not observed upon growth termination because the growth was carried out under N excess. However, for temperatures in the 500-585ºC range In droplets were observed. The reason is that, for temperatures above ~500ºC, InN decomposition provides excess In that partially reincorporates into the crystal until effective stoichiometry (at the growth front) is reached (growth starts under N-rich conditions). Beyond this point the excess In generates droplets at the surface that eventually disappear by desorption at high enough temperatures (>585 ºC) [7,13]. The growth is then limited by N, and the $[InN]^*$ value is lower than the nominal one by an amount that increases with the growth temperature [Fig 1.(a)]. These morphology changes are similar to that reported by Gallinat *et al.* for In-face InN [7].

As in InGaN alloys [9], we may expect that, under steady state conditions, InN losses ($\Phi_{InN}^{losses}$) by thermal decomposition are proportional to the Boltzmann factor with an activation energy, $E_a$, and to $[InN]^*$ with a constant factor $C$:

$$\Phi_{InN}^{losses}(T) = C \cdot [InN]^* \cdot \exp(-E_a / k_B T) \tag{1}$$

The nominal In incorporation rate, $\Phi_{In}^{inc}$, within the 500-585 ºC temperature range (In droplets region) when decomposition and desorption are not considered (ideal case), equals ($\Phi_N - \Phi_{Al}$) [9]. When decomposition and desorption are considered, an *actual* In incorporation rate can be defined, $\Phi_{In}^{inc\,*}$, which relates to the $[InN]^*$ value. The $\Phi_{InN}^{losses}$ were estimated within this temperature range as the $\Phi_{In}^{inc} - \Phi_{In}^{inc\,*}$ difference being the latter derived from HR-XRD data: $[InN]^* = \Phi_{In}^{inc\,*}/(\Phi_{In}^{inc\,*} + \Phi_{Al})$. Figure 1(b) shows an Arrhenius plot of $\Phi_{InN}^{losses}$ normalized to $[InN]^*$ values, that yield $C$ and $E_a$ values [best fit to eq.(1)] of $1.27 \times 10^{27}$ InN/s·cm$^2$ and 2.0 eV, respectively. This energy value is in good agreement with both the In-N bond energy, 1.93 eV [14], and the activation energy for thermal decomposition of In-face InN, 1.92 eV [7].

Figure 2 shows a growth diagram for metal-face InAlN determined from the temperature dependences of both $\Phi_{InN}^{losses}$ [eq.(1)] and In desorption rates from liquid In [7]. Notice that the diagram is also qualitatively valid for different values of the impinging metal and N fluxes. The horizontal dotted line, labelled as $\Phi_{In} = \Phi_N - \Phi_{Al}$, represents the nominal $\Phi_{In}$ value required to reach stoichiometric conditions for the given values of $\Phi_N$ and $\Phi_{Al}$, when neither In desorption, nor InN decomposition are considered. The selected value of $\Phi_{In}$ for Series-A is lower, indicating nominal N-rich conditions. Four different regimes are distinguished as a function of the impinging flux, $\Phi_{In}$, and the substrate temperature, namely: N-rich, In-droplets, intermediate In-rich, and no In, that is, where In incorporation is negligible. The solid line starting at low temperatures, that separates N-rich and In-droplets regimes, departs from the horizontal dotted line because excess In is available at the growth front due to InN decomposition. As temperature increases, more In excess is available, thus, less impinging $\Phi_{In}$ is required for stoichiometry conditions. Above this solid line, the In excess cannot be neither desorbed (below ~560 °C) nor incorporated to the crystal because of N shortage, thus, it accumulates on the surface as droplets. The stoichiometry condition determined by this solid line is given by:

$$\Phi_{In} + \Phi_{InN}^{losses}(T) = \Phi_N - \Phi_{Al} \qquad (2)$$

For temperatures above 560 °C, In desorption becomes non-negligible, opening a window in which growth can proceed under intermediate In-rich conditions [9,11,12]. As in the case of In(Al)GaN alloys the best quality is expected for those samples grown in this regime due to the presence of a surfactant In adlayer with coverages ranging from zero (lower N-rich boundary) to 2.5 monolayers (upper In-droplets boundary) [7, 9, 11-13]. Within this window (between the two solid lines) the required $\Phi_{In}$ increases with temperature to compensate In desorption. The lower boundary (towards N-rich) represents strict stoichiometry, whereas the upper one (towards In-droplets) is slightly metal rich. Both solid lines follow the expression:

$$\Phi_{In} + \Phi_{InN}^{losses}(T) - \Phi_{In}^{des}(T) = \Phi_N - \Phi_{Al} \qquad (3)$$

where $\Phi_{In}^{des}(T)$ reaches its maximum value (for a given temperature) at the upper boundary. These maximum values follow an Arrhenius temperature dependence with an activation energy of 2.49 eV [7]. The region where In incorporation is negligible, arises from InN-losses higher than the In incorporation rate limited by the impinging fluxes.

The diagram in figure 2 clearly shows that, for a given set of metals and N fluxes, the alloy composition changes significantly with the growth temperature. In addition, for a given temperature, the alloy composition is also expected to change with $\Phi_{In}$ outside the In-droplets regime. Figure 3 shows the increase of [InN]* as a function of $\Phi_{In}$ for Series-B. When stoichiometry is reached at the growth front (solid line boundary in figure 2) [InN]* reaches its maximum value, beyond which, In droplets develop while [InN]* remains constant.

Figure 4 shows the characteristic surface morphologies for samples grown at different regimes. Fig. 4(a) shows 1 - 2 μm size In droplets due to In accumulation (In-droplets regime). Figures 4 (b) and (c) show 5 x 5 μm$^2$ AFM images of samples grown in the intermediate In-rich and N-rich regimes, respectively. In the intermediate In-rich regime smooth surfaces with average roughness (rms) of 0.5 to 0.8 nm were observed. As shown in Fig. 4(b) the surface is characterized by atomic steps and spiral hillocks like those typical of the dislocation pinned step-flow growth of GaN layers grown by PA-MBE under Ga-rich conditions [see Fig. 4(d)]. Surprisingly, as shown in Fig. 4(c), for those samples grown in the N-rich regime smooth surfaces with atomic steps and rms values below 0.8 nm were obtained instead of the rough surfaces normally observed for III-N layers grown under N-rich conditions [5-8].

In summary, In incorporation into InAlN layers grown by PA-MBE was analysed as a function of the growth temperature and the impinging In flux. The In incorporation into the layers was found to decrease with the substrate temperature due to thermal InN decomposition and to increase with the impinging In flux until stoichiometry was reached at the growth front. From the temperature dependences of both InN losses and In desorption rates from liquid In a growth diagram was constructed showing four different growth regimes: N-rich, In-droplets, intermediate In-rich and a fourth one where In incorporation is negligible. This diagram provides a useful guide to control both alloy composition and surface morphology of InAlN layers grown by PA-MBE.

**Acknowledgments**

We acknowledge fruitful discussions with J. Pereiro, A. Bengoechea, J Grandal, M Utrera, and A. Redondo-Cubero. This work was partially supported by research grants from the Spanish Ministry of Education (NAN04/09109/C04/2, Consolider-CSD 2006-19); and the Community of Madrid (S-0505/ESP-0200).

.

**List of Figures**

thermodynamical limit for Ga-droplets formation.

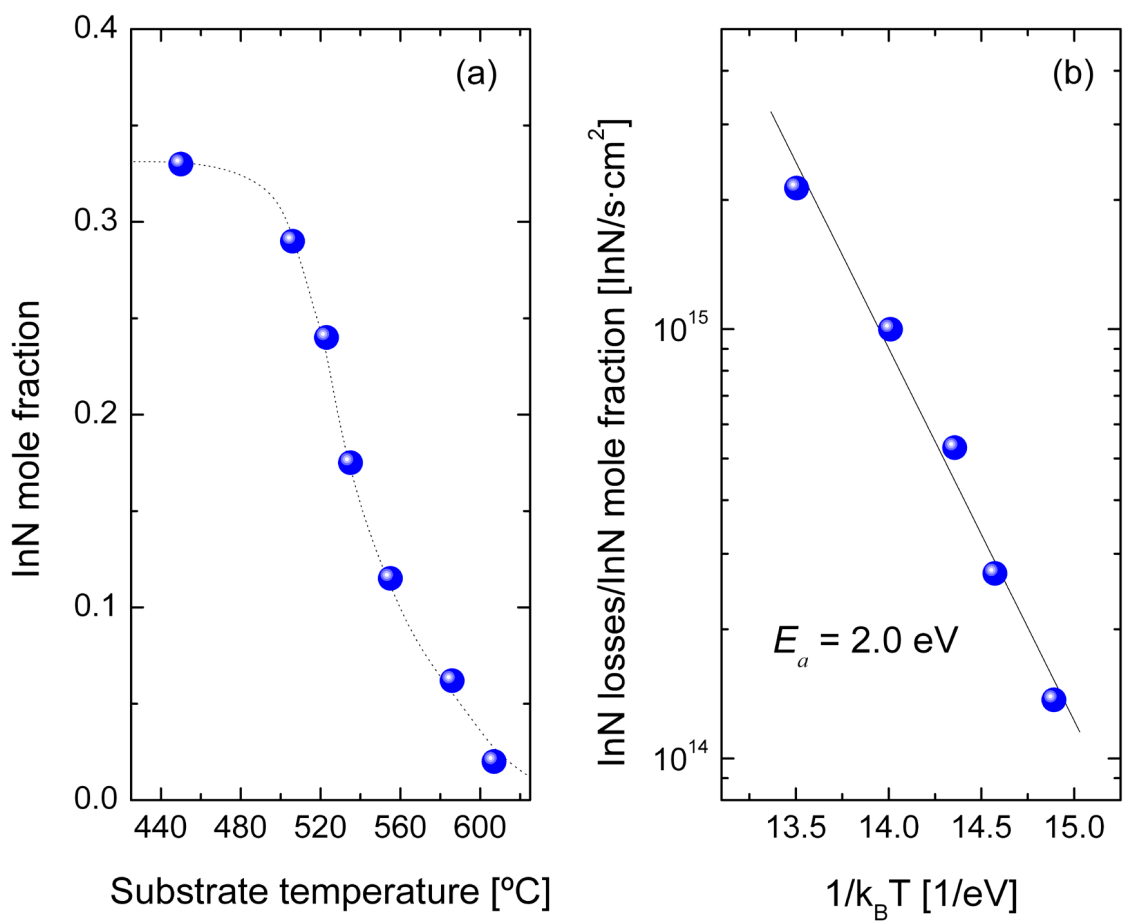

Figure 1

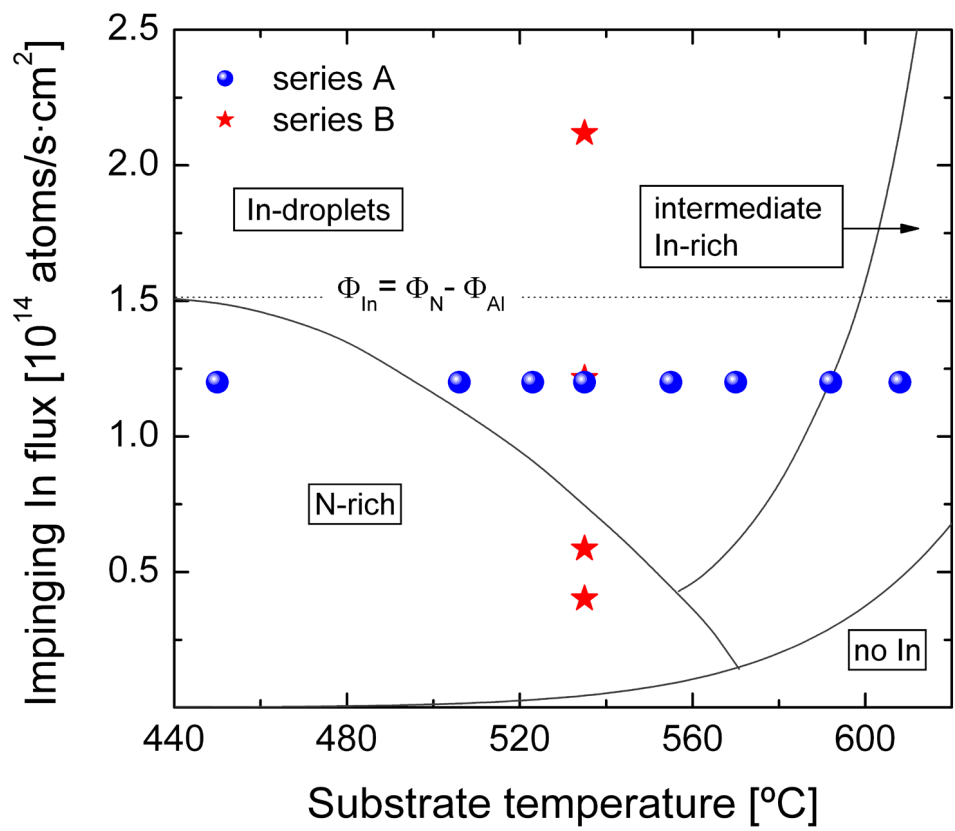

Figure 2

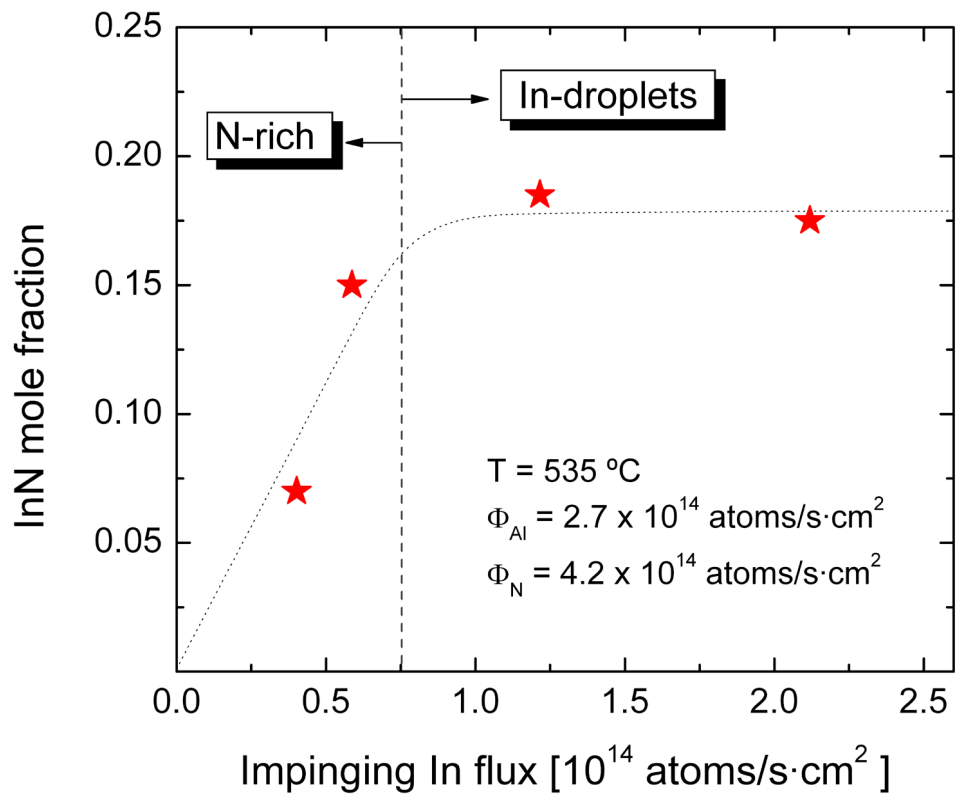

Figure 3

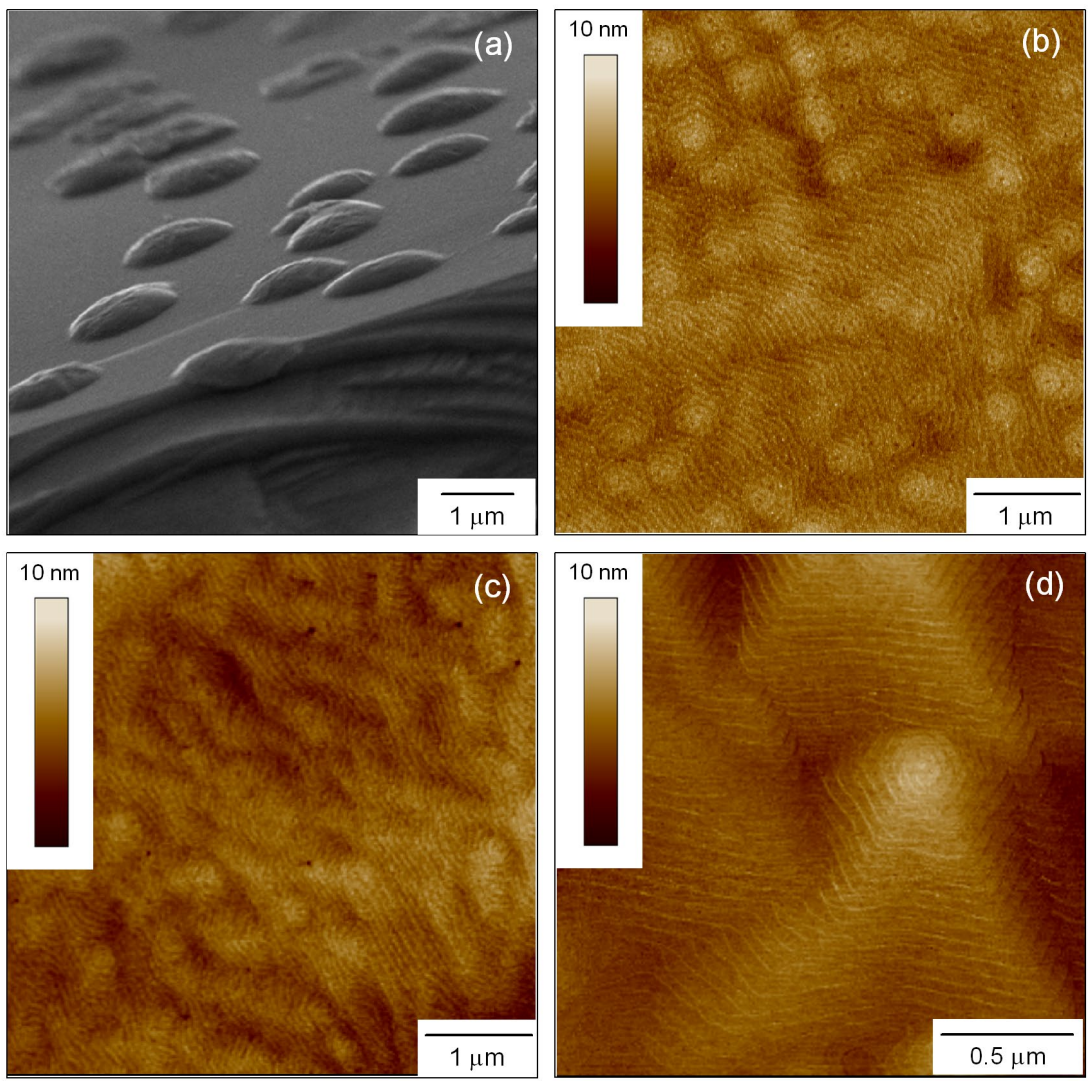

Figure 4